# Efficacy of *Wolbachia*-mediated sterility to suppress dengue: a synthetic control study

Jue Tao Lim, PhD[1,2,*], Somya Bansal, MSc[3,*], Chee Seng Chong, PhD [2,*], Borame Dickens, PhD[3], Youming Ng, BSc[2], Lu Deng, BSc[2], Caleb Lee, MSc[2], Li Yun Tan, BSc[2], Grace Chain, BSc[2], Pei Ma, MSc[3], Shuzhen Sim, PhD [2], Cheong Huat Tan, PhD[2], Alex R Cook, PhD[3], Lee Ching Ng, PhD [2,4,#]

[1]Lee Kong Chian School of Medicine, Nanyang Technological University, Singapore

[2]Environmental Health Institute, National Environment Agency, Singapore

[3]Saw Swee Hock School of Public Health, National University of Singapore and National University Health System, Singapore

[4]School of Biological Sciences, Nanyang Technological University, Singapore

[*]these authors contributed equally

[#]Correspondence to: Lee Ching Ng

Email: ng_lee_ching@nea.gov.sg

Environmental Health Institute, National Environment Agency, Singapore

11 Biopolis Way, Singapore 138667

## Abstract

**Background**

Solutions are needed to mitigate the spread of dengue due to the lack of available therapeutics and good vaccines. Matings between male *Aedes aegypti* mosquitoes infected with *w*AlbB strain of *Wolbachia* and wildtype females yield non-viable eggs. We evaluated the efficacy of releasing *w*AlbB-infected *Ae. aegypti* male mosquitoes to suppress dengue.

**Methods**

We conducted large-scale field trials in Singapore involving release of *w*AlbB-infected *Ae. aegypti* male mosquitoes for dengue control via vector population suppression, from epidemiological week (EW) 27 2018–EW 26 2022. All intervention and control locations practiced the same baseline dengue control protocol. To compare four towns which had releases progressively expanded versus controls, we used the synthetic control method to generate appropriate counterfactuals for intervention towns using a weighted combination of control towns (n=30) over 2014–2022. The main outcome was weekly dengue incidence rate caused by any dengue virus serotype.

**Findings**

Our study comprised an at-risk population of 607,872 individuals living in intervention sites and 3,894,544 individuals living in control sites. Interventions demonstrated up to 77·28% [121/156] (95% CI: 75·81–78·58) intervention efficacy despite incomplete coverage across all towns until EW26 2022. Intervention efficacies increased as release coverage improved over the years across all intervention sites. Releases led to 2,242 (95% CI 2,092–2,391) fewer cases per 100,000 persons in intervention sites during the study period. Secondary analysis showed that these intervention effects were replicated across all age groups and both sexes for intervention sites.

**Interpretation**

Our results demonstrated the potential of *Wolbachia*-mediated IIT for strengthening dengue control in tropical cities, where dengue burden is the greatest.

**Funding**

This study was supported by funding from Singapore's Ministry of Finance, Ministry of Sustainability and the Environment, National Environment Agency, and National Robotics Program. JTL is supported by the Ministry of Education (MOE), Singapore Start-up Grant. SB and ARC are supported by an MOE Tier 2 grant.

## Introduction

Dengue is the most widespread arboviral disease worldwide and showed sustained increases in burden over years. The Americas and Southeast Asia routinely account for the majority of global cases[1]. Singapore, a tropical, highly urbanized city state, is vulnerable to explosive dengue outbreaks due to a confluence of risk factors, including conducive conditions for year-round *Aedes* mosquito breeding; dense human population; and low population immunity due to decades of effective source reduction[2,3].

Conventional vector control remains the primary tool for mitigating the spread of dengue due to the lack of available therapeutics and good vaccines globally. Though these measures have successfully reduced the burden of dengue in Singapore[3,4], they yield diminishing returns as *Aedes aegypti* populations plunged. Against the backdrop of climate change and lowering herd immunity, conventional approaches for vector control are insufficient to mitigate dengue outbreaks. Therefore, there is a pertinent need for new vector control strategies.

Incompatible insect technique (IIT) is a promising complementary strategy for control of arbovirus transmission, involving releases of male mosquitoes infected with *Wolbachia*, a maternally inherited endosymbiotic bacterium. Due to cytoplasmic incompatibility[5,6], matings between *Wolbachia*-infected males and wildtype females not infected with the same *Wolbachia* strain yield non-viable eggs. Repeated releases of *Wolbachia*-infected males are thus expected to suppress wildtype mosquito populations and reduce disease transmission. Like classical sterile insect technique (SIT), where releases of irradiation-sterilized males have led to large-scale elimination of agricultural pests[7], IIT avoids disadvantages associated with traditional vector control, including genetic or behavioural resistance to insecticides and the inability to locate cryptic larval sites. IIT further avoids fitness costs arising from exposure to male-sterilizing doses of irradiation, which may reduce the mating competitiveness of released sterile males in an SIT program[8].

However, imperfect sex-sorting may lead to stable establishment of the released *Wolbachia* strain in the field due to unintentional release of fertile *Wolbachia*-infected female mosquitoes[9]. While this confers a reduced ability for the *Wolbachia*-established population to transmit dengue (a phenomenon exploited by some control programmes in an alternative *Wolbachia* strategy, where both male and female *Wolbachia*-infected mosquitoes are released to introgress the bacterium in the field population[10]), establishment renders cytoplasmic incompatibility-based IIT ineffective[9].

As part of efforts to augment Singapore's vector control programme with new tools, we have conducted extensive field trials of *Wolbachia*-mediated IIT targeting *Aedes aegypti*. To reduce the likelihood of stable establishment, we (**1**) combined IIT with SIT using low-dose irradiation to sterilize residual females during releases of *Wolbachia*-infected males or (**2**) a high-fidelity sex-sorting methodology in different intervention sites[11]. Here, we report the results of these extensive field trials, assessing the efficacy of deployments of male *Ae. aegypti* mosquitoes infected with the *w*AlbB strain of *Wolbachia* in reducing the incidence rate of dengue in Singapore as part of a suppression strategy.

**Research in context**

**Evidence before this study** It is unclear whether releases of *Wolbachia*-infected *Ae. aegypti* male mosquitoes are effective for the control of dengue. We searched for evaluations of interventions aimed at evaluating the effect of *Wolbachia*-infected *Ae. aegypti* male mosquitoes releases for the control of dengue. We searched EMBASE, MEDLINE, Global Health and Pubmed, up to August 15, 2023 (with no specified earliest date), with the search terms capturing the type of intervention ((("Wolbachia") OR ("incompatible insect technique")) and ("intervention")) as well as the type of outcome (dengue) OR (Aegypti) of interest for this study. We also contacted key field experts for relevant articles. The search returned 277 articles, of which 5 articles showed the efficacy of *Wolbachia* introgression approach on dengue incidence in Australia, Malaysia, Indonesia and Brazil. One article has examined the effect of incompatible insect technique on *Ae. aegypti* abundance. To the authors' knowledge, one article was not included in the search, which showed the effect of incompatible insect technique on *Ae. aegypti* abundance in Fresno County, California, United States.

This review highlighted a lack of robust evaluation on the effect of incompatible insect technique on dengue incidence.

**Added value of this study** This is the first analysis on the impacts of *Wolbachia*-infected *Ae. aegypti* male mosquito release on dengue incidence globally. Using a novel quasi-experimental approach, we identified that *Wolbachia*-infected *Ae. aegypti* male mosquito releases provide protection against dengue incidence across all large-scale field trial release sites.

**Implications of all the available evidence** Our evaluation provides robust support that *Wolbachia*-infected *Ae. aegypti* male mosquito releases can dramatically reduce dengue incidence. This evidence is critical given the urgent need for new classes of vector control to stem the global increase in dengue incidence.

**Methods**

**Deployment of male *Wolbachia*-infected *Ae. aegypti* and entomological monitoring** Releases were conducted twice weekly (weekdays, 0630–1030 hrs) at four designated public locations in high-rise public housing estates covering 607,872 individuals as of Epidemiological Week (EW) 26 2022. Bukit Batok, Choa Chu Kang and Yishun towns were subjected to interventions which combined IIT with SIT. Tampines town used the high-fidelity sex-sorting methodology and progressively adopted SIT protocols to release irradiated mosquitoes from January 2020. To trial whether *Aedes aegypti* population suppression could be sustained over increasingly larger areas, two large towns (Yishun, Tampines) were selected to adopt an expanding release strategy, where release sites were gradually expanded to adjacent neighbourhoods. Whereas Bukit Batok and

Choa Chu Kang towns were selected based on their smaller size to adopt a targeted release approach, which focused releases on areas with high *Aedes aegypti* abundance and persistent dengue transmission. (Table 1, Supplementary Information (SI) section 3).

| Township | Bukit Batok | Choa Chu Kang | Tampines | Yishun |
|---|---|---|---|---|
| Intervention start date | EW23 2020 | EW20 2020 | EW39 2018 | EW27 2018 |
| Study end date | EW26 2022 | EW26 2022 | EW26 2022 | EW26 2022 |
| Intervention time (weeks) | 109 | 112 | 197 | 209 |
| Total township size (m$^2$)$^\&$ | 627,441 | 1,145,559 | 5,088,046 | 3,473,690 |
| Average coverage over study period (%)* | 56.88% | 53.05% | 29.51% | 35.27% |
| Production Approach** | IIT-SIT | IIT-SIT | High fidelity sex-sorting | IIT-SIT |
| Frequency of release | Twice weekly | Twice weekly | Twice weekly | Twice weekly |
| Release strategy*** | Targeted$^\#$ | Targeted | Expanding$^{\#\#}$ | Expanding |
| Number of mosquitoes released | 6–7 w AlbB-SG males were released per study site resident per week | | | |
| Total number of mosquitoes released (rounded to thousands) | 17,139,000 | 17,139,000 | 17,139,000 | 17,139,000 |
| Township population over study period | 40,132 | 64,672 | 272,048 | 231,020 |

Table 1: Summary of *Wolbachia* intervention approaches over 4 townships.

$^\&$total area of public housing estates in respective townships

*Computed as (sum of area of releases multiplied by weeks of release till end of study period) over (total area of township multiplied by total release weeks). Areas were considered covered once they receive at least 6 months of Wolbachia interventions.

**The IIT-SIT approach and high-fidelity sex-sorting were detailed in supplementary information section 1 and has been previously characterised[13,16].

***denotes approach to releasing Wolbachia-infected males

$^\#$Targetted approach which focused releases on areas with high *Aedes aegypti* abundance and persistent dengue transmission

To ensure an even distribution of mosquitoes, releases were conducted at 6–12 equally spaced release locations per apartment block, with half of mosquitoes released in ground and the other half at upper floors alternating between middle (levels 5–6) and high floors (levels 10–11). Adult *Aedes aegypti* populations in release and control sites were monitored weekly using an average of six Gravitraps[12] per apartment block. Gravitraps were placed in public spaces along corridors and were evenly vertically distributed throughout the block, corresponding to a ratio of approximately one trap for every 20 households. Donor sites used to construct synthetic controls were all other towns which did not have *Wolbachia* releases up till EW26 2022, and comprise 30 towns with a population of 3,894,544 individuals. All intervention and donor townships in Singapore practiced the same vector control protocol before and after *Wolbachia* interventions took place[2,3] (SI section 3).

**Data** Dengue is a notifiable disease in Singapore and incidence data is collected by the Ministry of Health for all virologically confirmed cases. Cases have household addresses recorded and tagged to the town level for EW1 2014–EW26 2022 for analysis. Dengue cases are patients with virologically confirmed DENV infection through testing positive for RT-qPCR, NS1 or IgM. Clustered cases were taken as those which reside in a dengue cluster, which is defined as two or more cases with onset days within 14 days, located within 150 m of each other based on their house address, according to the operational criteria of the National Environment Agency, Singapore[13]. Sporadic cases are defined as those who do not reside in a dengue cluster. We normalized incidence at the town level by the number of Singapore residents in the release area of each town multiplied by 100,000 to obtain incidence per 100,000 persons (incidence rate) from EW1 2014–EW26 2022.

We also extracted a comprehensive set of spatial and spatio-temporally explicit variables to represent environmental heterogeneity across towns. Data sources and processing procedures were detailed in the SI Section 2.

**Estimation of intervention efficacy** We examined the impact of releasing male *Ae. aegypti* infected with *Wolbachia* on dengue incidence rates in four large-scale release sites (Bukit Batok, Choa Chu Kang, Yishun and Tampines towns) from EW1 2014 to EW26 2022 using the synthetic control method (SCM)[14]. The use of SCM is motivated by differences in pre-intervention trends in dengue incidence rates among intervention and non-intervention sites, which makes direct comparison between sites difficult. SCM alleviates this difficulty by generating an optimal set of weights for each control site using pre-intervention information to generate synthetic controls. This is done by minimizing the differences in weekly incidence rates between the pre-intervention control sites adjusted by the synthetic control weights with the intervention sites. The method re-weights control site locations in the study period such that the average pre-intervention trend of dengue incidence rates and covariates is similar to the intervention site of interest. The method first calculates the importance of considered covariates for dengue incidence rates and then subsequently compute weights which minimised the difference between both synthetic control and intervention sites in the importance-weighted covariates.

The intervention effect can then be estimated thereafter as the difference in incidence rates between intervention and synthetic control sites (the counterfactual). Intervention efficacy (IE) was defined as the percentage reduction in total dengue incidence rates per year (or across entire intervention period) in intervention sites with respect to the total dengue incidence rates in the synthetic control sites. Whereas number of cases averted was defined as the difference in the absolute number of dengue cases in intervention sites with respect to their synthetic control site for a pre-defined period. IEs were aggregated on the yearly basis to examine yearly changes in intervention efficacy due to inter-epidemic or epidemic phases of dengue transmission, and be interpreted in conjunction with year-on-year changes in *Wolbachia* coverage in each township. To construct 95% confidence intervals, we generated 100 bootstrap samples of dengue incidence rates using 'meboot' package in R. We employed the same fitting procedure to all bootstrapped timeseries to construct the empirical distribution of the synthetic controls where 95% confidence intervals can be obtained.

**Accounting for covariates** Inclusion of both the outcome variable of interest (dengue incidence rates) together with confounders in the optimization process of standard SCM would lead to confounders being ignored in the weighting process[15], potentially leading to a weighting scheme where covariates are not balanced between synthetic controls and intervention sites. This violates assumptions under the standard synthetic control method and may potentially bias intervention efficacy estimates. To obviate this risk, we explored three alternative estimators for the weighting scheme. Namely, using pre-intervention values of the dependent variable, as well as observed values of the covariates, we calculated SCM covariate weights using all pre-intervention covariates and dengue incidence rate observations (**M1**), the latest pre-intervention value of all covariates and incidence rates (**M2**), the pre-intervention time-average of all covariates and incidence rates (**M3**) and using all pre-intervention incidence observations without covariates (**M4**). **M1** and **4** represents the standard SCM with and without covariates respectively. **M2** is motivated by arguing that weighting based on the final value enables us to achieve a good fit at the pre-intervention cut-off time. **M3** is motivated by past work arguing that averaging both covariate and dependent variable values lead to constant weightage between both variable types in SCM weight estimation[16].

Covariates considered for balancing within the SCM weighting procedure included a high dimensional set of environmental/anthropogenic factors, where several were associated to vector abundance and dengue incidence in the study setting[17,18]. Covariate balance was assessed using mean differences or standardized mean differences. As **M1** failed to take into account any covariates, it was precluded from all other assessments (SI section 4). We examined the pre-intervention bias of these estimators using out-of-sample checks, and **M3** was the estimator which demonstrated the best predictive performance across most forecast horizons and provided good balance between covariates. It was thus taken as the estimator for the weighting scheme. SI section 4 and 5 provides full details of all weighting schemes, robustness checks and assessment approaches.

**Subgroup analysis on age and sex** Age-specific differences in the force-of-infection of dengue in Singapore have been reported [4]. Similarly, releases of male *Ae. aegypti* infected with *Wolbachia* may reduce dengue incidence rates in intervention sites, but dengue incidence rates may be mediated by immunity levels in each age group or sexes. Intervention efficacies may also be mediated by transmission patterns (clustered/non-clustered cases). Using the optimal weighting scheme, we re-estimated synthetic control weights for intervention sites using age- and sex-specific incidence rates. Intervention efficacy for each subgroup was then estimated by similarly computing the differences in synthetic control and intervention sites post-intervention.

This project was exempted from formal bioethics review as it is not considered human biological research, as advised by the Ministry of Health, Singapore. Only retrospective analysis using national dengue surveillance data was collected and approved by the Ministry of Health, Singapore, for this purpose. Permission to use anonymised dengue case data was obtained from the Ministry of Health, Singapore.

## Role of the funding source

The sponsor of the study had no role in study design, data collection, data analysis, data interpretation, or writing of the report. The corresponding author had full access to all the data in the study and had final responsibility for the decision to submit for publication.

## Results

Baseline characteristics of study population were presented in Table 2. The synthetic control arm had characteristics which were well-matched to the intervention group in the pre-intervention period.

| Characteristics | Control Group | | Intervention Group | | | | | | | | Synthetic Control Group | | | | | | | |
|---|---|---|---|---|---|---|---|---|---|---|---|---|---|---|---|---|---|---|
| Towns | | | Bukit Batok | | Choa Chu Kang | | Tampines | | Yishun | | Bukit Batok | | Choa Chu Kang | | Tampines | | Yishun | |
| Period | 2014 - 2018 | 2019 - 2022 | Pre-release | Post-release | Pre-release | Post-release | Pre-release | Post-release | Pre-release | Post-release | Pre-release | Post-release | Pre-release | Post-release | Pre-release | Post-release | Pre-release | Post-release |
| *Spatial-temporal characteristics* | | | | | | | | | | | | | | | | | | |
| Dengue Incidence Rate (per 100, 000 individuals) | 2·85 (6·94) | 8·31 (22·53) | 1·9 (2·91) | 2·79 (4·78) | 3·1 (5·32) | 3·29 (3·49) | 2·72 (2·81) | 3·26 (3·55) | 2·39 (2·46) | 2·39 (2·29) | 2·02 (1·64) | 6·19 (8·17) | 3·08 (2·99) | 8·62 (11·53) | 2·73 (2·51) | 5·77 (6·97) | 2·39 (2·03) | 6·66 (9·01) |
| Maximum Temp (ºC) | 35·33 (1·03) | 35·22 (1·08) | 34·41 (1·05) | 34·2 (1·03) | 34·49 (1·07) | 34·34 (1·07) | 34·38 (0·98) | 33·92 (0·98) | 34·69 (1·05) | 34·35 (1·08) | 34·4 (1·01) | 34·08 (1·01) | 34·34 (1) | 33·85 (1) | 34·25 (0·97) | 33·89 (1) | 34·62 (1·03) | 34·54 (1·06) |
| Mean Temp (ºC) | 27·61 (0·3) | 28·02 (0·84) | 27·68 (0·5) | 27·69 (0·81) | 27·65 (0·49) | 27·71 (0·82) | 27·77 (0·11) | 28·2 (0·79) | 27·55 (0·13) | 27·73 (0·82) | 27·73 (0·44) | 27·83 (0·81) | 27·8 (0·45) | 27·97 (0·81) | 27·65 (0·05) | 28·07 (0·77) | 27·56 (0·08) | 27·93 (0·78) |
| Minimum Temp (ºC) | 20·92 (0·95) | 22·07 (0·8) | 22·86 (0·81) | 23·46 (0·7) | 22·73 (0·78) | 23·38 (0·7) | 22·93 (0·9) | 23·5 (0·79) | 21·97 (0·7) | 22·02 (0·8) | 22·53 (0·83) | 23·61 (0·71) | 23 (0·86) | 23·83 (0·73) | 23·1 (0·89) | 23·34 (0·79) | 22·04 (0·8) | 23·33 (0·68) |
| Rainfall (mm) | 6·18 (5·37) | 6·66 (5·49) | 6·44 (5·05) | 8·28 (5·39) | 6·5 (5·28) | 8·37 (6·11) | 5·21 (5·02) | 6·21 (5·78) | 6·44 (5·84) | 6·8 (5·29) | 6·22 (5·17) | 8·03 (5·22) | 5·81 (5·1) | 7·76 (5·51) | 5·86 (5·01) | 6·64 (5·42) | 6·25 (4·95) | 6·77 (5·25) |
| Highest 30-min Rainfall (mm) | 3·04 (2·47) | 3·39 (2·62) | 3·32 (2·36) | 4·31 (2·59) | 3·71 (2·96) | 4·54 (3·21) | 2·6 (2·28) | 3·12 (2·67) | 2·87 (2·47) | 2·98 (2·48) | 3·12 (2·27) | 4·12 (2·46) | 2·98 (2·26) | 3·81 (2·46) | 3·09 (2·34) | 3·42 (2·52) | 2·93 (2·14) | 3·42 (2·47) |
| Highest 60-min Rainfall (mm) | 3·74 (3·16) | 4·21 (3·41) | 4·09 (3) | 5·39 (3·42) | 4·53 (3·7) | 5·69 (4·29) | 3·19 (2·93) | 3·92 (3·53) | 3·54 (3·16) | 3·67 (3·16) | 3·85 (2·92) | 5·15 (3·21) | 3·67 (2·9) | 4·76 (3·22) | 3·79 (2·99) | 4·26 (3·28) | 3·59 (2·7) | 4·23 (3·21) |
| Highest 120-min Rainfall (mm) | 4·2 (3·63) | 4·8 (3·99) | 4·59 (3·46) | 6·19 (4·08) | 5·1 (4·2) | 6·49 (5·03) | 3·57 (3·38) | 4·46 (4·14) | 3·93 (3·54) | 4·22 (3·74) | 4·33 (3·4) | 5·92 (3·79) | 4·12 (3·36) | 5·44 (3·77) | 4·25 (3·44) | 4·84 (3·82) | 4·02 (3·09) | 4·8 (3·75) |
| Mean Wind Speed | 7·9 (2·12) | 8·78 (2·37) | 7·2 (1·82) | 8·24 (1·26) | 7·08 (1·9) | 8·62 (1·35) | 9·4 (2·38) | 9·39 (2·45) | 7·64 (2·26) | 8·15 (2·38) | 8·07 (1·89) | 7·71 (1·38) | 8·24 (2·05) | 8·28 (1·63) | 8·09 (1·86) | 8·62 (1·85) | 7·64 (1·83) | 8·56 (2·33) |
| Max Wind Speed | 33·56 (3·85) | 35·07 (4·18) | 31·67 (3·26) | 32·37 (2·54) | 31·93 (3·45) | 32·74 (2·56) | 33·88 (3·36) | 34·76 (3·49) | 33·39 (4·08) | 33·92 (3·37) | 34·49 (3·64) | 35·24 (2·98) | 34·23 (3·52) | 35·43 (3·17) | 34·43 (3·31) | 35·97 (3·54) | 33·52 (3·36) | 34·87 (3·34) |
| Mean Relative Humidity | 80·04 (3·05) | 79·7 (3·23) | 79·79 (3·37) | 79·85 (3·03) | 79·78 (3·38) | 79·89 (3) | 79·99 (2·9) | 79·87 (3·3) | 80·09 (2·93) | 79·77 (3·24) | 79·83 (3·26) | 79·97 (2·88) | 79·84 (3·21) | 80·06 (2·78) | 79·98 (2·91) | 79·86 (3·3) | 80·09 (2·93) | 79·76 (3·24) |
| *Spatial Characteristics* | | | | | | | | | | | | | | | | | | |
| NDVI (Vegetation Index) | 0·32 (0·04) | | 0·35 (0) | | 0·28 (0) | | 0·33 (0) | | 0·33 (0) | | 0·33 (0) | | 0·3 (0) | | 0·31 (0) | | 0·33 (0) | |
| Area within 300m of a waterbody (%) | 0·32 (0·18) | | 0·35 (0) | | 0·34 (0) | | 0·24 (0) | | 0·16 (0) | | 0·39 (0) | | 0·34 (0) | | 0·49 (0) | | 0·32 (0) | |
| Area within 500m of a waterbody (%) | 0·55 (0·23) | | 0·7 (0) | | 0·68 (0) | | 0·49 (0) | | 0·43 (0) | | 0·67 (0) | | 0·59 (0) | | 0·7 (0) | | 0·58 (0) | |
| Vegetation Density | 0·05 (0·06) | | 0·02 (0) | | 0·02 (0) | | 0·04 (0) | | 0·01 (0) | | 0·05 (0) | | 0·03 (0) | | 0·05 (0) | | 0·04 (0) | |

| Characteristic | | | | | | | | | |
|---|---|---|---|---|---|---|---|---|---|
| Average public housing building height (m) | 22·33 (16·12) | 26·38 (0) | 36·19 (0) | 30·75 (0) | 31·41 (0) | 35·07 (0) | 36·19 (0) | 31·79 (0) | 22·37 (0) |
| Average age of public housing (years) | 17·51 (12·02) | 36·01 (0) | 31·43 (0) | 29·23 (0) | 30·99 (0) | 27·23 (0) | 26·33 (0) | 26·64 (0) | 21·29 (0) |
| Average public housing price (SGD) | 786943·87 (1133104·7) | 4785332·5 (0) | 4785332·5 (0) | 426132·2 (0) | 2967732 (0) | 1055575·47 (0) | 759965·86 (0) | 1088115·58 (0) | 515951·26 (0) |
| Distance of centroid to drainage network (m) | 399·22 (178·48) | 145·16 (0) | 345·73 (0) | 435·87 (0) | 469·72 (0) | 304·89 (0) | 345·74 (0) | 322·99 (0) | 380·98 (0) |
| Length of drainage network inside spatial unit (m) | 71271·75 (2961·95) | 1725·55 (0) | 833·68 (0) | 3699·05 (0) | 2341·02 (0) | 2434·48 (0) | 2580·25 (0) | 3692·14 (0) | 1068·84 (0) |
| Forest area (%) | 0·01 (0·03) | 0 (0) | 0 (0) | 0 (0) | 0 (0) | 0·01 (0) | 0 (0) | 0·01 (0) | 0 (0) |
| Grass area (%) | 0·01 (0·02) | 0 (0) | 0 (0) | 0 (0) | 0 (0) | 0·01 (0) | 0·01 (0) | 0·01 (0) | 0·03 (0) |
| Total vegetation area (%) | 0·02 (0·01) | 0·01 (0) | 0·02 (0) | 0·02 (0) | 0·01 (0) | 0·02 (0) | 0·02 (0) | 0·02 (0) | 0·02 (0) |
| Building area (%) | 0·28 (0·06) | 0·27 (0) | 0·23 (0) | 0·26 (0) | 0·27 (0) | 0·27 (0) | 0·3 (0) | 0·29 (0) | 0·27 (0) |
| Number of condo units | 4010 (176·59) | 0 (0) | 0 (0) | 0 (0) | 2 (0) | 85·1 (0) | 113·88 (0) | 168·07 (0) | 125·72 (0) |
| Number of landed units | 60941 (2951·6) | 0 (0) | 0 (0) | 0 (0) | 0 (0) | 1366·69 (0) | 921·54 (0) | 2159·61 (0) | 2237·55 (0) |
| Number of public housing units | 874393 (23074·84) | 46230 (0) | 49026 (0) | 75342 (0) | 67908 (0) | 40738·99 (0) | 33525·64 (0) | 39345·77 (0) | 33748·7 (0) |

Table 2: Baseline characteristics of study population pre- and post-*Wolbachia* releases in intervention, control and synthetic control group. The pre-release period for each intervention town is from start of study till last pre-intervention week. The post-release period for each intervention town is from intervention start week to end of study period. The numbers in bracket represent standard deviation for each characteristic.
[1]Maximum (minimum) temperature and windspeed were calculated by taking maximum (minimum) of temperature and windspeed across all towns within intervention or control groups. Length of drainage network, number of condo units, number of landed units and number of public housing units were calculate by taking sum across all towns within intervention or control groups. The remaining characteristics were calculated by averaging across all towns within intervention or control groups. All the calculations were done for the specified time period.
[2]The characteristics for synthetic control group were first estimated by multiplying characteristics with weights from synthetic control method produced by best model diagnostics (**M3**). The summary statistics for each summary statistic was determined in the same way as intervention and control groups.

**Suppression of wild-type *Ae. aegypti* populations in intervention sites** Suppression of adult wild-type *Ae. aegypti* populations was demonstrated across all trial sites, with the Gravitrap *Aedes aegypti* Index (GAI) progressively reduced as *Wolbachia* coverage increased geographically. When >50% coverage was achieved by 2022, the overall GAI of all sites plunged below 0.05 (SI section 3). 80% suppressive efficacy on *Ae. aegypti* abundance was achieved when areas experienced more than 6 months of *Wolbachia* releases (SI section 7).

**Intervention efficacy of IIT for four field trial sites** *Wolbachia* interventions were associated with significantly decreased in annual dengue incidence rates across all trial sites versus synthetic controls (Figure 1). Comparing each intervention site's yearly dengue incidence rates versus their respective synthetic controls and actual controls constituting the synthetic control, demonstrated that yearly dengue incidence rates were higher in 83.8% [90/107 town-years] of all constituent controls versus intervention sites and higher in all synthetic control sites versus intervention sites in the post-intervention period (Figure 1).

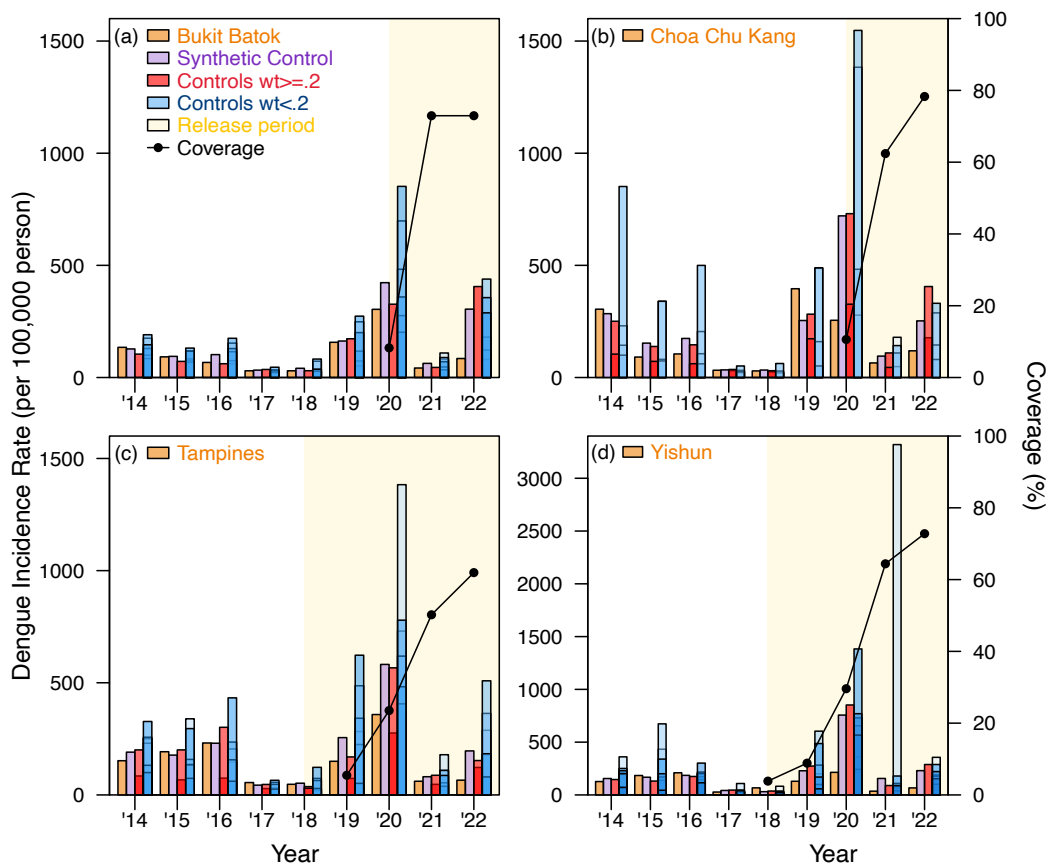

[**Figure 1**: Dengue incidence rate from EW1 2014 to EW26 2022 in the intervention sites of (a) Bukit Batok, (b) Choa Chu Kang, (c) Tampines and (d) Yishun. Figure shows the dengue incidence per 100,000 by year in public housing areas in each intervention town (orange) and its corresponding synthetic control town (purple). The dengue incidence in each control town which were selected to be constituents of the synthetic control were also superimposed in the same bar, and visualized separately by those which constituted high (wt ≥ 0.2) and low weightages (wt ≤ 0.2) respectively within each town's respective actual synthetic control. Higher weightages denote locations with more contribution on the synthetic control's dengue incidence rate.

The geographical coverage (%) represents the percentage of areas within the town which is covered by *Wolbachia* interventions for at least six months and is calculated at the end of each year. Points represent coverage of *Wolbachia* interventions by the end of each year. The six month mark for coverage is based on the time it takes *Wolbachia* release to have around 80% suppressive efficacy on *Ae. aegypti* abundance (SI section 7).]

Intervention efficacy (IE) and number of cases averted due to *Wolbachia* interventions for a pre-defined period was compared against the geographical coverage (%) of *Wolbachia*. Geographical coverage represented the percentage of areas within the town which was covered by *Wolbachia* interventions for at least six months until that year. The six month mark for coverage was based on the time it takes for *Wolbachia* interventions to have around 80% suppressive efficacy on *Ae. aegypti* abundance in the designated release site (SI section 7).

Yearly IEs in the four trial sites, and aggregate yearly IEs increased concomitantly with increased *Wolbachia* intervention coverage (Table 3,4). In 2019, at around 5·67% coverage, aggregate IE was around 42·32% [205/484] (95% CI: 40·01%–44·46%). IEs increased (57·65% [1197/2077], 95% CI: 56·23%–58·98%) as coverage increased to 23·30% in 2020, with an almost four-fold increase in number of cases averted in 2020 (2114·63, 95% CI: 1998·79–2230·46) versus 2019 (516·90, 95% CI: 470·66–563·13). This may be due to the expansion of *Wolbachia* to two additional sites in 2020. A slight dip in intervention efficacies in 2021 (48·52% [192/396], 95% CI: 45·73%–51·03%) may be due to 2021 being an inter-epidemic year, with a far lower number of cases averted compared to 2020 despite the increased coverage in 2021. Upon achievement of 68.08% coverage in EW26 2022, the highest level of IE was achieved at 65·81% [647/984] (95% CI: 64·24%–67·26%), with individual town IEs ranging from 52·85% to 72·21% for that year. In EW1–EW26 of 2022, *Wolbachia* interventions were associated to 906·88 (95% CI: 850·94–962·82) cases being averted on aggregate.

| Year | | Bukit Batok | Choa Chu Kang | Tampines | Yishun | All Sites |
|---|---|---|---|---|---|---|
| 2019 EW1-52 | Total cases IE (%) | – | – | 41·30 [106/256] (39·49 – 43·01) | 43·46 [99/229] (40·60 – 46·05) | 42·32 [205/484] (40·01 – 44·46) |
| | Clustered cases IE (%) | – | – | 50·23 [104/208] (47·72 – 52·52) | 69·61 [181/261] (68·28 – 70·84) | 61·02 [286/468] (59·20 – 62·69) |
| | Sporadic cases IE (%) | – | – | 9·12 [5/51] (7·27 – 10·91) | -0·6 [-0·3/50] (-3·83 – 2·43) | 4·33 [4/101] (1·83 – 6·71) |
| | Total cases averted* | – | – | 287·10 (266·31 – 307·90) | 229·79 (204·35 – 255·24) | 516·90 (470·66 – 563·13) |
| | *Wolbachia* coverage | 0·00% | 0·00% | 5·46% | 8·85% | 5·67% |
| 2020 EW1-52 | Total cases IE (%) | 39·20 [104/266] (37·12 – 41·15) | 69·21 [327/472] (68·01 – 70·32) | 38·40 [223/582] (36·73 – 39·98) | 71·73 [543/757] (70·74 – 72·66) | 57·65 [1197/2077] (56·23 – 58·98) |
| | Clustered cases IE (%) | 10·80 [16/148] (-0·49 – 20·35) | 67·93 [219/323] (64·61 – 70·7) | 42·63 [230/540] (38·57 – 46·41) | 80·25 [651/811] (79·61 – 80·85) | 61·27 [1116/1822] (58·83 – 63·51) |
| | Sporadic cases IE (%) | 20·2 [8/37] (15·78 – 24·17) | -22·72 [-8/34] (-28·64 – -17·32) | 23·97 [15/64] (21·85 – 25·97) | 10·2 [6/60] (4·3 – 15·69) | 10·88 [21/195] (6·6 – 14·88) |
| | Total cases averted* | 41·91 (38·37 – 45·44) | 211·28 (199·86 – 222·7) | 607·77 (566·08 – 649·45) | 1253·68 (1194·49 – 1312·88) | 2114·63 (1998·79 – 2230·46) |
| | *Wolbachia* coverage | 8·26% | 10·60% | 23·52% | 29·60% | 23·30% |
| 2021 EW1-52 | Total cases IE (%) | 32·87 [21/63] (29·87 – 35·63) | 31·94 [30/95] (28·18 – 35·33) | 24·81 [20/81] (21·80 – 27·59) | 77·28 [121/156] (75·81 – 78·58) | 48·52 [192/396] (45·73 – 51·03) |
| | Clustered cases IE (%) | 29·63 [5/18] (-1·58 – 49·44) | 46·97 [23/50] (38·12 – 55·08) | 37·67 [15/40] (29·18 – 47·34) | 78·18 [40/52] (72·06 – 82·23) | 52·88 [84/158] (42·51 – 61·39) |
| | Sporadic cases IE (%) | 24·02 [9/39] (21·27 – 26·58) | -25·81 [-8/31] (-31·8 – -20·34) | 8·22 [3/40] (5·68 – 10·62) | 21·54 [7/31] (18·05 – 24·74) | 8·13 [11/140] (4·64 – 11·38) |
| | Total cases averted* | 8·32 (7·24 – 9·41) | 19·71 (16·48 – 22·94) | 54·76 (46·27 – 63·26) | 278·87 (256·95 – 300·78) | 361·67 (326·95 – 396·39) |
| | *Wolbachia* coverage | 72·95% | 62·40% | 50·20% | 64·40% | 57·70% |
| 2022 EW1-26 | Total cases IE (%) | 72·21 [220/305] (70·91 – 73·39) | 52·85 [133/253] (50·43 – 55·05) | 66·48 [130/196] (65·47 – 67·43) | 71·01 [163/230] (69·47 – 72·41) | 65·81 [647/984] (64·24 – 67·26) |
| | Clustered cases IE (%) | 65·74 [115/175] (63·18 – 68·21) | 65·62 [159/243] (62·80 – 68·21) | 77·71 [120/155] (75·61 – 79·81) | 82·90 [176/213] (81·98 – 83·74) | 72·71 [571/785] (70·67 – 74·67) |
| | Sporadic cases IE (%) | -4·06 [-1/24] (-17·72 – 6·94) | -59·18 [-13/22] (-71·31 – -48·65) | -22·46 [-6/26] (-27·41 – -17·88) | -15·67 [-4/26] (-23·87 – -8·5) | -24·52 [-24/97] (-34·22 – -16·07) |
| | Total cases averted* | 88·32 (82·88 – 93·77) | 86·32 (78·34 – 94·31) | 354·93 (339·32 – 370·54) | 377·3 (350·39 – 404·20) | 906·88 (850·94 – 962·82) |
| | *Wolbachia* coverage | 72·95% | 78·31 | 61·94% | 72·79% | 68·07% |
| Across intervention period | Total cases IE (%) | 54·43 [345/634] (52·55 - 56·16)*** | 59·84 [491/820] (58·04 – 61·48) | 43·02 [480/1115] (41·34 – 44·6)*** | 67·53 [926/1372] (66·1 – 68·84) | 56·88 [2242/3941] (55·18 – 58·46) |
| | Clustered cases IE (%) | 39·96 [136/340] (33·25 - 46·03)*** | 65·33 [402/615] (61·87 – 68·38) | 49·87 [470/942] (46·23 – 53·4)*** | 78·51 [1049/1336] (77·53 – 79·43)*** | 63·60 [2057/3234] (61·04 – 66) |

Table 3: Intervention efficacy (IE) of *Wolbachia* releases on total and clustered dengue incidence rates across four trial sites. IE is calculated after start of intervention for all sites, hence 2019 "All sites" IE does not include Bukit Batok and Choa Chu Kang, while 2020 "All site" IE includes Bukit Batok and Choa Chu Kang only after their release started. – not estimable as *Wolbachia* releases did not start. Total dengue cases comprise the sum of sporadic and clustered dengue cases. *case aversion at four *Wolbachia* field trial sites by comparing dengue incidence in counterfactual versus dengue incidence in *Wolbachia* trial sites. Numbers

in parenthesis represent upper and lower bounds for 95% confidence intervals. Fraction in square brackets represtent change in dengue incidence rate per 100,000/counterfactual dengue incidence rate per 100,000. *Wolbachia* coverage computed as (area within specific site covered by *Wolbachia* interventions times number of effective months)/(final coverage of *Wolbachia* interventions in site times number of effective months). Areas were considered covered once they receive at least six months of *Wolbachia* interventions. ***indicates significant differences at the 0.05 level in post-intervention treatment effect for all trial sites compared to the pre-intervention period according to Andrews' test for **M3** across all intervention time.

Taken together, *Wolbachia* releases were estimated to have an aggregate IE of 56·88% [2242/3941](95% CI: 55·18%–58·46%) and 63·60% [2057/3234](95% CI: 61·04%–66.00%) over EW1 2019–EW26 2022 for total and clustered cases, respectively (Table 4) in the four field trial sites, despite only having an aggregate coverage of 34·49% in all sites during the field trial period.

| Subgroup | Intervention Efficacy (%)* | | | | |
|---|---|---|---|---|---|
| | 2019 | 2020 | 2021 | 2022 | Aggregate |
| Total | 42·32 [205/484] (40·01 – 44·46) | 57·65 [1197/2077] (56·23 – 58·98) | 48·52 [192/396] (45·73 – 51·03) | 65·81 [647/984] (64·24 – 67·26) | 56·88 [2242/3941] (55·18 – 58·46) |
| Coverage[#] | 5·67% | 23·30% | 57·70% | 68·07% | 34.49%[##] |
| Subgroup by dengue case type | | | | | |
| Clustered | 61·02 [286/468] (59·20 – 62·69) | 61·27 [1116/1822] (58·83 – 63·51) | 52·88 [84/158] (42·51 – 61·39) | 72·71 [571/785] (70·67 – 74·67) | 63·60 [2057/3234] (61·04 – 66) |
| Sporadic | 4·33 [4/101] (1·83 – 6·71) | 10·88 [21/195] (6·6 – 14·88) | 8·13 [11/140] (4·64 – 11·38) | -24·51 [-24/97] (-34·22 – -16·07) | 2·43 [13/535] (-2·1 – 6.62) |
| Subgroup by sex | | | | | |
| Female | 51·38 [133/260] (49·56 – 53·08) | 56·02 [463/826] (54·00 – 57·87) | 37·52 [51/135] (30·70 – 43·39) | 72·06 [287/398] (70·39 – 73·58) | 57·67 [934/1619] (55·43 – 59·72) |
| Male | 41·88 [110/262] (39·74 – 43·87) | 50·71 [519/1024] (49·12 – 52·2) | 41·51 [85/204] (37·67 – 44·9) | 63·80 [374/586] (62·26 – 65·21) | 52·38 [1087/2075] (50·53 – 54·10) |
| Subgroup by age | | | | | |
| Age 0 – 6 | -19·16 [-1/3] (-32·06 – -4·34) | 38·68 [6/14] (24·84 – 51·57) | 72·41 [2/3] (68·11 – 80·37) | 62·38 [8/13] (56·01 – 68·26) | 44·95 [15/33] (34·75 – 55·36) |
| Age 7 – 20 | 66·24 [53/80] (64·34 – 68·38) | 62·58 [149/238] (60·28 – 64·63) | 58·55 [14/24] (42·82 – 82·36) | 85·63 [116/136] (84·07 – 87·17) | 69·54 [333/478] (66·92 – 73·32) |
| Age 21 – 60 | 42·63 [143/335] (40·77 – 44·38) | 55·92 [705/1261] (54·40 – 57·33) | 52·92 [125/236] (49·50 – 55·92) | 66·78 [498/746] (65·64 – 67·84) | 57·06 [1471/2579] (55·45 – 58·56) |
| Age 61+ | 41·11 [39/95] (38·74 – 43·30) | 22·63 [66/292] (13·90 – 31·40) | -10·30 [-8/74] (-23·94 – 6·84) | 67·99 [135/199] (65·81 – 69·97) | 35·27 [233/661] (29·49 – 41·29) |

Table 4: Aggregate intervention efficacy of *Wolbachia* releases on total and clustered dengue incidence rates as well as subgroups across all trial sites. Total dengue cases comprise the sum of sporadic and clustered dengue cases.

*intervention efficacy estimates at four *Wolbachia* trial sites by comparing dengue incidence rates in synthetic controls which never experienced released versus dengue incidence rates in *Wolbachia* trial sites. Numbers in parenthesis represent upper and lower bounds for 95% confidence intervals. Fraction in square brackets represent change in dengue incidence rate per 100,000/counterfactual dengue incidence rate per 100,000

[#]*Wolbachia* coverage computed as (area within specific site covered by *Wolbachia* interventions times number of effective weeks)/(total area of *Wolbachia* interventions in site times number of weeks in a year). An area within the town is considered covered by *Wolbachia* interventions if it has experienced release for at least six months until that year.

[##]Coverage across all towns for all years from 2019 to EW26 2022 calculated same as above with area as total area across sites covered by *Wolbachia* intervention.

We also note that *Wolbachia* releases had larger IEs on clustered dengue incidence rate vs sporadic dengue incidence rate (Table 3, SI section 6), suggesting that the latter may have lowered the aggregate intervention efficacy based on total dengue cases. There were no noticeable changes in the direction and significance of aggregate IEs across both sexes (Table 4), but estimated effect sizes per year varied considerably (SI section 6). Larger IEs were also found in adolescents (7–20) and adults (21–60), vs elderly (61+) and children (0–6) (Table 4, SI section 6). This is likely due to incidence rates being highest in adolescents and adults in the study setting (SI sections 6).

**Role of the funding source** The *Wolbachia* programme was funded by the Ministry of Finance. The funders of this study had no role in the study design, data collection and writing of this report. Data analysis was conducted at Nanyang Technological University and the National University of Singapore.

**Discussion**

Despite incomplete coverage over sites over the study period, releases of *w*AlbB-infected *Ae. aegypti* male mosquitoes were associated with a dramatic reduction in dengue incidence rates (Table 3, 4). Intervention efficacies increased concomitantly with coverage (Table 4, SI section 6). On aggregate, across all towns and years, we estimated a 56·88% [2242/3941](95% CI: 51·88%–58·46%) reduction in dengue incidence rate at an average coverage of 34·49%, and a 65·81% [647/984](95% CI: 64·24%–67·26%) reduction in dengue incidence rate in EW1–EW26 2022 when coverage was at 68·07% (Table 4). In sites where coverage was high, and interventions were employed over a long timeframe, such as Yishun, we estimated a 71·01% [163/230](95% CI: 69·47%–72·41%) reduction in dengue incidence rates at 72·79% coverage in 2022 (Table 3). This protective efficacy was demonstrated in all four field trial sites, across sexes and age groups, over epidemic and inter-epidemic years. The demonstration of protective effects across all four trial sites, subgroups and years support consistent biologic replication of the intervention effects (Table 4). Releases of *w*AlbB-infected *Ae. aegypti* male mosquitoes at sustained 100% coverage per intervention site would potentially maximize intervention efficacy estimates, given the upward trajectory in intervention efficacies demonstrated till EW26 2022 as coverage increased. While full coverage of interventions in these four field trial sites was achieved post EW26 2022, the initiation of a large-scale cluster-randomized control trial comprising an additional 15 intervention and control locations made it difficult to construct an appropriate donor pool for further evaluation[19].

Efficacy results reported here are consistent with previous laboratory and entomological field observations. Sterile insect technique combined with incompatible insect technique does not significantly affect fitness cost in released mosquitoes[20]. Release of incompatible *Ae. aegypti* male mosquitoes drives profound suppression of wild-type mosquitoes[9,11,21,22]. While previous field trials[10] have demonstrated the efficacy of *w*Mel introgression in reducing dengue incidence, no study has yet examined the effect of incompatible insect technique for *Ae. aegypti* on reducing dengue incidence. Our study combines data from large-scale field trial releases and utilized a robust, novel quasi-experimental framework to demonstrate the protective efficacy of *w*AlbB-infected *Ae. aegypti* release on dengue.

Incompatible insect technique using *w*AlbB-infected *Ae. aegypti* male mosquitoes represents a new class of tools for the control of dengue. This strategy has several advantages, (**1**) while protective efficacy is only demonstrated for dengue in this study, as it is the only *Aedes*-borne disease in constant circulation in the study sites, this efficacy should be similar against other *Ae. aegypti*-borne diseases as it suppresses vector populations through cytoplasmic incompatibility rather than blocking disease transmission under the introgression approach. (**2**) High public acceptance towards the technology has been previously demonstrated[23], and (**3**) it has been retrospectively shown that the technology can be cost-effective in reducing dengue incidence in the study setting at the 40% efficacy threshold[24]. The threshold was met at the aggregated 65·81% protective efficacy in 2022 (Table 2). As parts of each site had only received treatment for less than 6 months, efficacy can improve with longer periods of releases. Wider release areas can potentially also address sporadic cases by reducing number of "imports" from non-release areas, and improve efficacy. (**4**) IIT/SIT technologies have demonstrated scalability in larger and/or less resource-rich settings[25,26] (**5**) Lastly, while dengue virus could plausibly evolve resistance to *Wolbachia* under the introgression approach[27], IIT suffers no drawbacks related to *Wolbachia*-associated selective pressure of viruses.

However, several limitations do exist. Our study relies on non-randomized but large-scale field trial data to assess protective efficacy on dengue incidence. We relied on quasi-experimental approaches which can appropriately account for the observational nature of data, and considered a large set of environmental and anthropogenic factors which may confound intervention efficacy estimates. A cluster-randomized controlled trial is also underway to supplement these results and ascertain the utility of the intervention[19]. The current intervention is employed in high-rise public housing estates, which demonstrated the protective efficacy in urban household setting. Future work should look at the multivalency of the intervention in other study settings. As releases of *Wolbachia*-infected male *Ae. aegypti* is a biological intervention, the long-run efficacy needs to be determined, as fitness cost of *Wolbachia*-infected male *Ae. aegypti* may vary. Furthermore, the current analysis took the earliest start date of release in each town as the start-point of intervention. This downwardly biased IE estimates as the entire township is taken as the treated region rather the specific locale of intervention (SI section 3). Current analyses assumed no spillover effects of interventions into control sites. While care has been taken to ensure that natural boundaries such as roads and non-residential zones demarcate each intervention site (SI section 3), spillover effects were not accounted for in the analytical framework. If the donor pool was subject to spillover decreases in dengue incidence, these reductions would downwardly bias the synthetic control counterfactual and makes IE estimates conservative. Identification of the locations where dengue was acquired for each case is also difficult, as it is impossible to contact-trace dengue transmission chains. The use of nationally comprehensive dengue surveillance databases can alleviate potential reporting errors, together with estimating intervention efficacy by clustered and sporadic case definitions to account for potential importation of cases into intervention towns. The consistently low or negative intervention efficacies estimated for sporadic cases suggests that a proportion of cases may have acquired the disease elsewhere and biased the estimated total intervention efficacies downwards further.

As programs incorporating the release of *Wolbachia*-infected male *Ae. aegypti* scale up for future deployment, it should not be viewed as a complete replacement for conventional vector control methods. Public health authorities would do well to continue to set aside sufficient resources and capacity for continued source reduction efforts. In our experience, both the entomological and epidemiological impacts of the intervention are likely to be maximized if it is used to complement and enhance, rather than to replace, conventional vector control measures.

**Data sharing statement**

All code required to replicate study output will be made available with publication on github.com/somya-b/wolbachia-epi. The evaluation of the effect on dengue incidence was a retrospective analysis using national dengue surveillance data collected by the Ministry of Health. The databases used in this study are a property of the Ministry of Health, Singapore which were shared with the researchers under the legal mandate of the Infectious Disease Act, precluding data sharing with a third party.

**Acknowledgements**

The authors thank Prof. Zhiyong Xi of Michigan State University for providing the *w*AlbB-infected *Aedes aegypti* line, which was used to generate the *w*AlbB-Sg line essential to this study.

**Author Contributions**

Conceptualization: JTL, SB, CCS, CHT, NLC, Data Curation: BD, YN, LD, CL, LYT, GC, Formal Analysis: JTL, SB, Funding acquisition: NLC, Investigation: JTL, SB, BD, YN, LD, CL, LYT,GC, NLC, Methodology: JTL, SB, BD, YN, LD, CL, LYT,GC, NLC, Project administration: CCS, CHT, NLC, Resources: JTL, ARC, NLC, Software: SB, Supervision: JTL, ARC, NLC , Validation: JTL, SB, Visualization: JTL, SB, LYT, GC, Writing – original draft : JTL, SB, Writing – review & editing: All authors have read and reviewed the manuscript. JTL, SB, GC, LYT accessed and verified the data. JTL and NLC were responsible for the decision to submit the manuscript.

**Conflict of Interests**

The authors declare no conflict of interests